\documentclass[%
 preprint,
superscriptaddress,
 amsmath,amssymb,
 aps,
 prb,
]{revtex4-2}

\usepackage{physics}
\usepackage{braket}
\usepackage{qcircuit}
\usepackage{graphicx}
\usepackage{dcolumn}
\usepackage{bm}
\usepackage[hidelinks]{hyperref}
\usepackage{tensor}
\usepackage{xcolor}
\usepackage{threeparttable}
\usepackage{chemformula}
\usepackage[overload]{textcase}

\newcommand{\wtli}[1]{\textcolor{black}{#1}}
\newcommand{\wtliwtli}[1]{\textcolor{black}{#1}}

\begin{document}


\title{Quantum Machine Learning of Molecular Energies with Hybrid Quantum-Neural Wavefunction}
\newcommand{\CUHK}{School of Science and Engineering, The Chinese University of Hong Kong, Shenzhen, Guangdong 518172, China}
\newcommand{\THU}{Department of Chemistry, MOE Key Laboratory for Organic OptoElectronics and Molecular Engineering, Tsinghua University, Beijing 100084, China}
\newcommand{\IOP}{Institute of Physics, Chinese Academy of Sciences, Beijing 100190, China}

\author{Weitang Li}
\email{liwt31@gmail.com}
\affiliation{\CUHK}

\author{Shi-Xin  Zhang}
\affiliation{\IOP}

\author{Zirui Sheng}
\affiliation{\CUHK}

\author{Cunxi Gong}
\affiliation{\CUHK}

\author{Jianpeng Chen}
\affiliation{\CUHK}

\author{Zhigang Shuai}
\affiliation{\CUHK}
\affiliation{\THU}

\date{\today}

\begin{abstract}
Quantum computational chemistry holds great promise for simulating molecular systems more efficiently than classical methods by leveraging quantum bits to represent molecular wavefunctions. However, current implementations face significant limitations in accuracy due to hardware noise and algorithmic constraints. To overcome these challenges, we introduce a hybrid framework that learns molecular wavefunction using a combination of an efficient quantum circuit and a neural network. Numerical benchmarking on molecular systems shows that our hybrid quantum-neural wavefunction approach achieves near-chemical accuracy, comparable to advanced quantum and classical techniques. 
Based on the isomerization reaction of cyclobutadiene, a challenging multi-reference model, our approach is further validated on a superconducting quantum computer with high accuracy and significant resilience to noise.
\end{abstract}


\maketitle

\section{Introduction}
Quantum computers leverage quantum effects to store and manipulate data, making them particularly suitable for the simulation of microscopic quantum systems~\cite{daley2022practical,  king2024computational, chan2024quantum}. 
The Variational Quantum Eigensolver (VQE) algorithm is the most widely adopted framework for 
quantum computational chemistry~\cite{peruzzo2014variational, google2020hartree, cerezo2021variational, tilly2022variational, huang2022variational, guo2024experimental}.
The key component of the VQE algorithm is the parameterized quantum circuit, 
which learns the quantum state of the system under study variationally~\cite{yuan2019theory}.
The challenge of VQE lies in striking a delicate balance between circuit depth and accuracy~\cite{lee2018generalized, grimsley2019adaptive, sun2024toward, xiao2024physics}.
While deeper circuits tend to improve accuracy,
they also make the algorithm more sensitive to noise and and can suffer from barren plateaus~\cite{mcclean2018barren}.
In contrast, shallow circuits may not capture the system's complexity adequately.
Parallel to the evolution of VQE, 
Neural Networks (NNs) have shown remarkable success in representing quantum wavefunctions of chemical systems~\cite{hermann2023ab}.
Based on variational Monte Carlo, these NNs are trained to minimize the energy expectation, similar to the VQE approach.
Efforts along this line include DeepWF~\cite{han2019solving}, FermiNet~\cite{pfau2020ab}, 
PauliNet~\cite{hermann2020deep}, QiankunNet~\cite{shang2023solving}, and so on~\cite{li2022ab, scherbela2024towards, li2024improved, nys2024ab}.
Thanks to the expressive power of NNs, these methods demonstrate accuracy comparable to Coupled Cluster with Single and Double excitations (CCSD) but with significantly lower computational scaling, typically  $\mathcal{O}(N^4)$. 

The success in these new wavefunction representations has inspired the development of hybrid quantum-neural wavefunctions, where quantum circuits and neural networks are jointly trained to represent the wavefunction of quantum systems~\cite{zhang2022variational}.
In this hybrid approach, quantum circuits are responsible for learning the quantum phase structure of the target state,
which is a difficult task for neural networks alone~\cite{westerhout2020generalization},
and the neural network correctly describes the amplitude.
The combination of quantum computation and variational Monte Carlo has also demonstrated considerable potential in simulating quantum systems~\cite{jiang2024walking},
and the inclusion of neural networks significantly enhances the expressiveness of trial wave functions,
thereby leading to more accurate and scalable simulations.
The intersection between quantum computing and machine learning, known as quantum machine learning,
is developing at a rapid pace~\cite{biamonte2017quantum, cerezo2022challenges, ren2022experimental, li2024high}.
Chemistry applications include the construction of shallow depth ansatz,  energy eigenstate filtration, material phase prediction, neural network pertaining, and so on~\cite{li2021quantum, sajjan2021quantum, sajjan2022quantum, zeng2023quantum, halder2023machine, halder2024machine, shang2024rapidly}.

In this work, we propose a quantum machine learning framework for efficient representation of molecular wavefunction
and accurate computation of molecular energies. 
The method employs the linear-depth paired Unitary Coupled-Cluster (UCC) with Double excitations (pUCCD) circuit to learn molecular wavefunction in the seniority-zero subspace~\cite{henderson2015pair, elfving2021simulating, o2023purification, zhao2023orbital, khan2023chemically}, 
and a neural network to correctly account for the contributions from unpaired configurations.
We propose an efficient algorithm to compute the expectations of physical observables for the hybrid quantum-neural wavefunction,
which avoids calculating the overlap between the quantum circuit state and classical state, or the costly process of quantum state tomography. 
This represents an enhancement of scalability over the previously proposed quantum-classical hybrid quantum Monte-Carlo method~\cite{huggins2022unbiasing}.
We name our method as pUNN, which stands for paired Unitary coupled-cluster with Neural Networks.
pUNN retains the low qubit count ($N$ qubits) and shallow circuit depth of pUCCD, while achieving accuracy comparable to the most precise quantum and classical computational chemistry methods, such as UCCSD (UCC with single and double excitations) and CCSD(T) (CCSD with perturbative triple excitations). We demonstrate the efficacy of pUNN through numerical simulations of various diatomic and polyatomic molecular systems, such as \ch{N2} and \ch{CH4}. 
To test pUNN in a real quantum computing scenario, we compute the reaction barrier for the isomerization of cyclobutadiene on a programmable superconducting quantum computer. The results demonstrate that the pUNN is highly accurate and noise resilient
for a real quantum computing task.

\section{Theory and Methodology}

\begin{figure}[tbh]
\includegraphics[width=\linewidth]{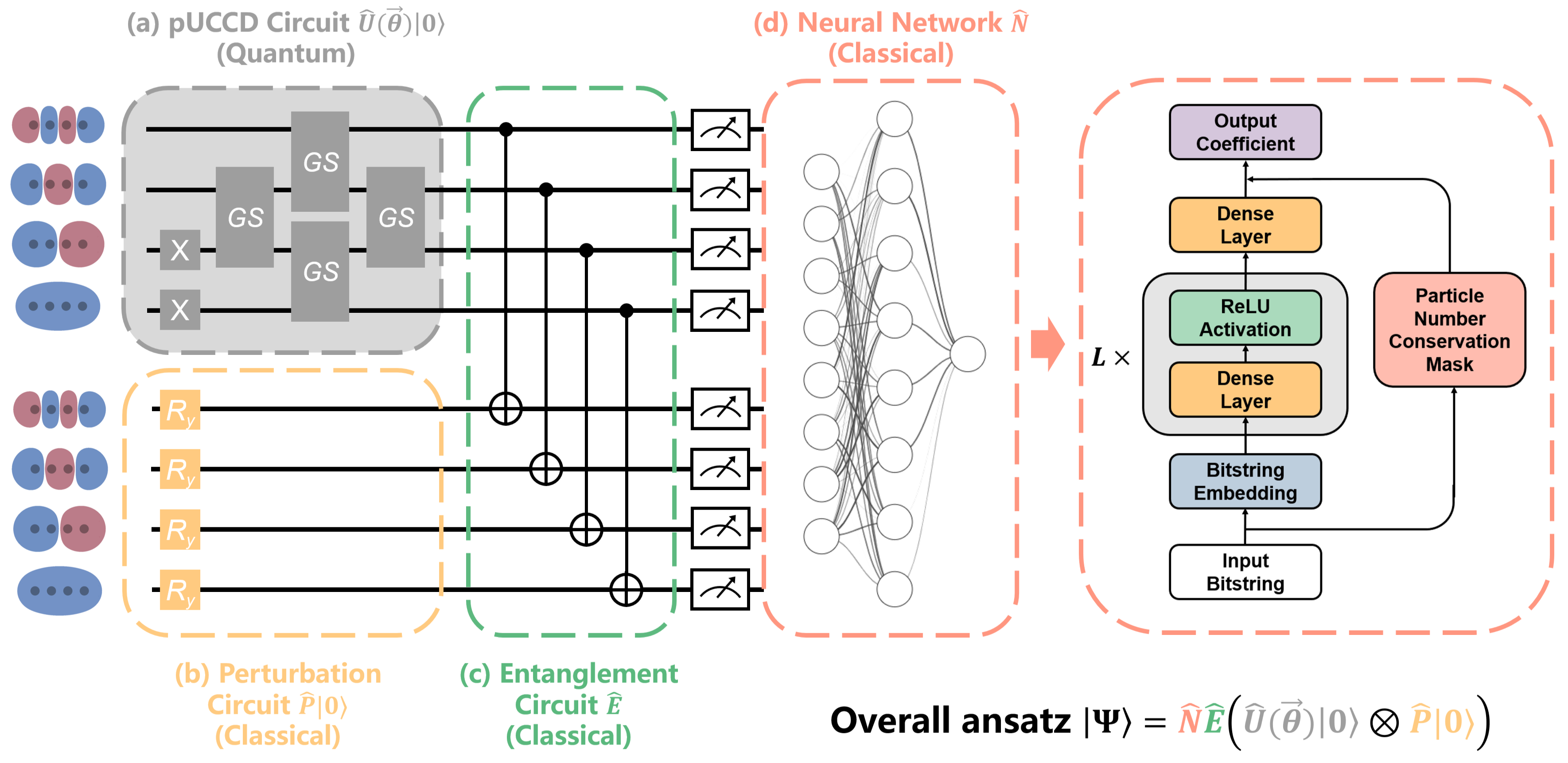}
\caption{\label{fig:diagram} 
A Schematic diagram for the pUNN framework. 
The pUCCD circuit in the grey box is the only component executed on a real quantum computer.
``GS'' denotes Givens-Swap gate.
Meanwhile, the perturbation circuit and the entanglement circuit are processed classically.
Together, the quantum circuit and the neural network serves as an ansatz 
and are trained jointly to represent the molecular wavefunction.
}
\end{figure}

In this section, we present our pUNN algorithm and focus on our contribution. 
General backgrounds, such as the electronic structure problem and the UCC types of ansatz
for quantum computational chemistry
are briefly overviewed in the Appendix A.
We start by employing the pUCCD ansatz to represent molecular wavefunction, 
which is encoded in the parameterized quantum circuit $\hat U(\vec \theta)$.
In the computational basis, the pUCCD circuit state can be expressed as
\begin{equation}
\label{eq:puccd_orig}
    \ket{\psi}=\sum_k a_k \ket{k} \ ,
\end{equation}
where $\ket{k}$ represents the occupation of a pair of electrons in the original $N$-qubit Hilbert space.
For ground state problems, the coefficients $a_k$ can be assumed to be real numbers.
To correctly describe the configurations outside of the seniority-zero subspace, we add $N$ ancilla qubits to the circuit and expand the Hilbert space from $N$ qubits to $2N$ qubits.
In the expanded $2N$-qubit space, the equivalent state is
\begin{equation}
\label{eq:puccd_expand}
    \ket{\Phi} = \sum_k a_k \ket{k}\otimes \ket{k} \ ,
\end{equation}
with the two $\ket{k}$ terms now representing the occupation of the alpha and beta spin sectors, respectively. 
We note that these $N$ ancilla qubits can be treated classically, which will be explained later.

In the context of quantum circuits, the expanded state $\ket{\Phi}$ is constructed from $\ket{\psi}$ using the ancilla qubits and an entanglement circuit $\hat E$:
\begin{equation}
    \ket{\Phi} = \hat E \left (\ket{\psi} \otimes \ket{0} \right) \ .
\end{equation}
The entanglement circuit $\hat{E}$ creates the necessary correlations between the original qubits and the ancilla qubits. $\hat E$ can be decomposed into $N$ parallel CNOT gates:
\begin{equation}
    \hat E = \prod_i^N \textrm{CNOT}_{i, i+N} \ ,
\end{equation}
where each CNOT gate entangles the $i$-th original qubit with the corresponding $i$-th ancilla qubit.

Although $\ket{\Phi}$ has $2N$ qubits while $\ket{\psi}$ has $N$ qubits, from a quantum chemistry perspective, they represent the same state in the seniority-zero space and therefore have the same energy. 
We then apply the neural network, acting as an quantum operator $\hat N$, on the quantum state.
$\hat N$ is a non-unitary post-processing operator~\cite{zhang2022variational} defined in the expanded Hilbert space.
After applying $\hat N$, the overall state becomes 
$\hat N \hat E \left ( \ket{\psi}\otimes \ket{0}\right )$.
The method is inspired by variational quantum-neural hybrid eigensolver (VQNHE) and it provides exponential acceleration for nonunitary postprocessing in VQE than naive transformed Hamiltonian approach \cite{Mazzola2019_z, Benfenati2021a_z, Shang2023_z}.
The neural network operator $\hat N$ modulates the state $\ket{\Phi}$ as follows:
\begin{equation}
\label{eq:def-nn}
    \hat N = \sum_{kj} b_{kj} \ket{k}\ket{j}\bra{j}\bra{k} \ ,
\end{equation}
where $b_{kj}$ is a real tensor
 represented by a continuous neural network $\mathcal{B}(k, j)$, such that $b_{kj} = \mathcal{B}(k, j)$.
To drive $\hat N \hat E \left ( \ket{\psi}\otimes \ket{0}\right )$ out of the seniority-zero subspace, 
we  apply a perturbation circuit $\hat P$ to the 
ancilla qubits at the beginning, diverting the state of the ancilla qubits $\ket{\phi} = \hat P \ket{0}$ from $\ket{0}$
\begin{equation}
   \ket{\phi} = \hat P \ket{0} = \frac{ \ket{0} + \sum_{j\neq 0} \epsilon_j \ket{j} }{
    1 + \sum_{j\neq 0}\epsilon^2_j } \ ,
\end{equation}
where $\epsilon_j$ are small coefficients satisfying $\sum_{j\neq 0}\epsilon_j^2 \ll 1$.
As a result, our algorithm is expected to be resilient to noise \cite{Zhang2021d_z}, making it well-suited for implementation on real quantum devices. 
The conservation of the particle number is enforced by the neural network introduced in the following.
The values of $\epsilon_j$ and the exact form of $\hat P$ are flexible.
The only key requirement for $\hat P$ is that it should have a low circuit depth, which allows efficient simulation of $\hat P \ket{0}$ on classical computers.
To this end, we adopt a perturbation circuit with single qubit rotation gates $R_y$ for each qubit and the rotation angle is set to 0.2.
$\hat P$ produces real coefficients, a desired property for the ground state of the molecular Hamiltonian.

After describing the quantum circuit part, we turn to the neural network structure used for $\mathcal{B}(k, j)$.
$\mathcal{B}(k, j)$ accepts the two bitstring $k$ and $j$ as input and outputs the coefficients $b_{kj}$.
The first component of the neural network is
embedding the bitstring $\ket{k}\otimes \ket{j}$ into a vector. 
We employ a binary representation, where $\ket{k}\otimes \ket{j}$ is converted to a vector of size $2N$, with each element being either -1 or 1.
The vector $\mathbf{x}_{0}(k, j)$ is then passed through a neural network consisting of $L$ dense layers and ReLU activation functions
\begin{equation}
    \mathbf{x}_{i+1}(k, j) = \textrm{ReLU}\left [\mathbf{W}_i \mathbf{x}_i (k, j) + \mathbf{c}_i \right ] \ .
\end{equation}
In the hidden layers, the number of neurons is set to $2KN$ where $K$ is a tunable integer that controls the size of the neural network.
In this work we set $K=2$ unless otherwise specified.
The number of layers $L$ is set to $N-3$, proportional to the size of the molecule.
The number of parameters in the neural network scales
as $K^2N^3$ considering both the width and depth of the neural network.
The computational complexity is also $\mathcal{O}(K^2N^3)$ for each input bitstring.

The final dense layer outputs the desired coefficient $b_{kj}$, before multiplying with the particle number conservation mask $m(k, j)$
\begin{equation}
b_{kj} = m(k, j) \left [\mathbf{W}_L \mathbf{x}_L (k, j) + \mathbf{c}_L \right] \ .
\end{equation}
The mask $m(k,j)$ is defined as
\begin{equation}
m(k,j) = \begin{cases}
\mathbf{1} & \textrm{if} \quad \sum_i k_i = N_{\alpha} \quad \textrm{and} \quad \sum_i j_i = N_{\beta}, \\
\mathbf{0} & \textrm{otherwise},
\end{cases}
\end{equation}
where $N_{\alpha/\beta}$ is the number of spin up/down electrons.
The mask eliminates configurations $\ket{k}\otimes \ket{j}$ that do not conserve the number of spin up and down electrons.

To summarize, the overall wavefunction is given by
\begin{equation}
\label{eq:overall-ansatz}
    \ket{\Psi} = \hat N \hat E \left ( \hat U(\vec \theta) \ket{0} \otimes \hat P \ket{0} \right ) \ ,
\end{equation}
which consists of four components: the pUCCD circuit $\hat U(\vec \theta)$, the perturbation circuit $\hat P$, the entanglement circuit $\hat E$ and the neural network $\hat N$.
The next challenge is to measure the expectation value of 
the physical observables such as the energy based on Eq.~\eqref{eq:overall-ansatz},
which is highly nontrivial without resorting to quantum tomography or incurring exponential measurement overhead. 
Without an efficient measurement protocol, the pUNN approach could be rendered impractical.
\wtli{
Besides, in quantum computational chemistry, the number of measurements required to estimate expectation values is a key indicator of efficiency for variational algorithms like pUNN. }
In fact, the ansatz represented by Eq.~\eqref{eq:overall-ansatz} is carefully designed in such a way 
that an efficient algorithm for computing expectation values is possible.

Since $\ket{\Psi}$ is not normalized, the energy expectation is
\begin{equation}
    \braket{E} = \frac{\braket{\Psi|\hat H|\Psi}}{\braket{\Psi|\Psi}} \ .
\end{equation}
Here we outline the key points of the measurement protocol that enables the computation of both $\braket{\Psi|\hat H|\Psi}$ and $\braket{\Psi|\Psi}$ using the measurement outcome of the quantum circuit $\hat U(\vec \theta)\ket{0}$ 
and the output from the neural network.
The full measurement protocol is provided in the Appendix B.
For brevity, we assume there is only a single Pauli string in $\hat H$, and the summation over many Pauli strings can be handled straightforwardly.
We also note that the estimation of the norm $\braket{\Psi|\Psi}$ can be considered as a special case when $\hat H = \hat I$.

To perform the measurement, we transform the Hamiltonian $\hat H$ and the neural network $\hat N$ with $\hat E$
\begin{equation}
\label{eq:exp1}
    \braket{\Psi|\hat H|\Psi} = \braket{\psi \otimes \phi|\left ( \hat E^\dagger \hat N^\dagger \hat E \right )
    \left ( \hat E^\dagger \hat H \hat E \right ) 
    \left ( \hat E^\dagger \hat N \hat E \right )
    |\psi \otimes \phi} \ .
\end{equation}
Since $\hat E$ is a Clifford circuit, $ \hat H' = \hat E^\dagger \hat H \hat E$ is also a Pauli string.
Additionally, since $\hat E$ is composed of CNOT gates, it reversibly maps one bitstring to another,
rather than a linear combination of bitstrings. Specifically, 
\begin{equation}
    \hat E \left ( \ket{k}\otimes \ket{j} \right ) = \ket{k}\otimes \ket{k\oplus j} \ .
\end{equation}
The transformed neural network $\hat N' = \hat E^\dagger \hat N \hat E$ is
\begin{equation}
    \hat N' = \sum_{kj} b_{kj} \ket{k}\ket{k \oplus j}\bra{k \oplus j}\bra{k}
    = \sum_{kj} b_{k,k\oplus j} \ket{k}\ket{j}\bra{j}\bra{k} \ .
\end{equation}
$\hat N'$ is thus formally the same as $\hat N$ but with a permuted index for the coefficient $b$.

After the transformation, the entanglement circuit $\hat E$ is removed from Eq.~\eqref{eq:exp1}
\begin{equation}
\label{eq:two-circuit-measure}
    \braket{\Psi|\hat H|\Psi} = \braket{\psi \otimes \phi|\hat N'^\dagger \hat H' \hat N'
    |\psi \otimes \phi} \ .
\end{equation}
Eq.~\eqref{eq:two-circuit-measure} corresponds to the measurement of $\hat N'^\dagger \hat H' \hat N'$ on two unentangled circuits $\ket{\psi}$ and $\ket{\phi}$.
If $\hat N'$ is absent or if $\hat N' = \hat I$, the measurement of $\hat H$ can be performed efficiently by measuring the two separate circuits $\ket{\psi}$ and $\ket{\phi}$.
In Appendix B, we show that, by carefully designing the measurement circuit,
$\hat N' \hat H' \hat N'$ can also be measured by 
separate measurement of $\ket{\psi}$ and $\ket{\phi}$, with a constant overhead.
Therefore, the evaluation of $\braket{\Psi|\hat H|\Psi}$ is cast into the separate measurement of $\ket{\psi}$ and $\ket{\phi}$. 
Since $\ket{\phi}$ is designed to be a shallow circuit that can be efficiently simulated classically, 
the only circuit that needs to be executed on real quantum devices is the pUCCD circuit $\ket{\psi}$.
Nonetheless, the number of terms to measure in the Hamiltonian increases from $N^2$ in the pUCCD circuit to $N^4$ for more general electronic structure problems.
\wtli{Thus, in terms of measurement shots,
the pUNN method is as efficient as other quantum computational methods such as UCCSD , 
but with significantly reduced circuit depth and higher accuracy.}
Compared with Entanglement Forging~\cite{eddins2022doubling}, which utilizes classical sampling to recover the entanglement between two sub-systems,
our method encodes the entanglement between two sub-systems into $\hat H$ and $\hat N$ and avoids excessive sampling.

A schematic diagram of the pUNN framework is depicted in Fig.~\ref{fig:diagram}.
In the whole algorithm, only the pUCCD circuit within the grey box in dashed lines is executed on quantum computers,
which allows pUNN to maintain the $N$-qubit requirement for the computation instead of $2N$.
The perturbation circuit and entanglement circuit can be efficiently processed on classical computers.
The measured bitstring of the composite circuit is fed into the neural network for $\mathcal{B}(k, j)$,
which is then used to adjust the measurement outcome.
The entire ansatz is then set up in a VQE workflow, where both the parameters in the quantum circuit and the neural network are trained to minimize the molecular energy. This process ultimately yields the ground state through the variational principle.

In the noiseless simulation described in Sec.\ref{sec:noiseless}, we use the L-BFGS-B algorithm to optimize the parameters in the quantum circuit. For circuit optimization on real quantum hardware, we employ the SOAP method \cite{li2024efficient}.
For both the noiseless simulations and the experiments on quantum computers, the neural network is trained using the AdaMax optimizer~\cite{kingma2014adam}, a variant of the widely adopted Adam optimizer. 
The optimizer begins with a learning rate schedule of $\alpha = 0.01$, $b_1=0.8$ and $b_2=0.99$.
The learning rate decays linearly to $\alpha=0.001$ between the 8000th and 32000th steps.
This learning rate schedule helps ensure stable convergence by gradually decreasing the learning rate as the training progresses.
For noiseless simulation, the maximum number of steps is set to 64000.
\wtli{
A summary table for the hyper-parameters can be found in the Supplementary Information.
}
For the noiseless simulation, we initialize the neural network with five different random seeds, and the lowest energy found across these seeds is reported.
For quantum circuit manipulation, including both noiseless and noisy emulation as well as interfacing with real quantum hardware, we use the TensorCircuit framework~\cite{zhang2023tensorcircuit}. 
General quantum computational chemistry tasks, including Hamiltonian construction, reference value calculation, and parameter optimization are handled by TenCirChem~\cite{li2023tencirchem}, a specialized package built on top of TensorCircuit designed for quantum computational chemistry.
TenCirChem also relies on PySCF for evaluating the integrals and performing calculations based on classical computational chemistry~\cite{sun2020recent}.

\section{Results}
\subsection{Accuracy and Scalability}
\label{sec:noiseless}
\begin{figure}[tbh]
\includegraphics[width=\linewidth]{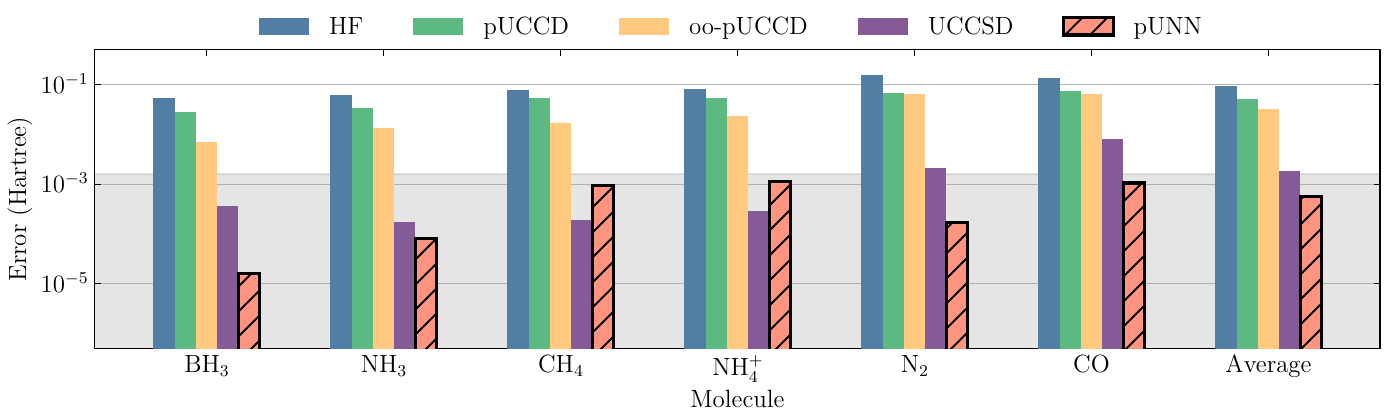}
\caption{\label{fig:error-quantum} 
Compare the accuracy of pUNN with other quantum computational chemistry methods.
The $1s$ orbitals are frozen and the reference energy is FCI.
The shaded area indicates the chemical accuracy.
}
\end{figure}

We first compare the accuracy of pUNN with other quantum computational methods in Fig.~\ref{fig:error-quantum}.
For this comparison, 
we perform noiseless numerical calculations on molecular systems corresponding to 8 spatial orbitals and 16 qubits. 
The basis set employed is STO-3G and the $1s$ orbitals are frozen.
The exact geometries of the molecules are reported in the Supplementary Information.
The full configuration interaction (FCI) energy for these molecules is computed as the reference energy.
As shown in Fig.~\ref{fig:error-quantum}, the standard pUCCD approach improves over the HF method but consistently shows the highest error across all molecules.
The results indicate that the neglect of configurations outside of the seniority-zero subspace limits the accuracy.
The orbital optimization pUCCD (oo-pUCCD) method~\cite{sokolov2020quantum, mizukami2020orbital, zhao2023orbital} reduces the error to a modest extent for most molecules,
except for \ch{N2} and \ch{CO}, where the errors of pUCCD and oo-pUCCD are comparable.
This demonstrates the limitation of oo-pUCCD, as it still assumes electron paring.
The UCCSD method, known for its high accuracy, performs well across the board.
However, UCCSD requires $2N$ qubits for $N$ molecular orbitals and has a very deep circuit, which is computationally expensive. 
The typical circuit depth of UCCSD for these molecules is approximately 1000.
Our proposed pUNN method stands out as the most accurate approach for the majority of the molecules studied. 
In the meantime, pUNN uses only $N$ qubits for $N$ molecular orbitals,
and its circuit depth is the same as the circuit depth of pUCCD.
In contrast to UCCSD, the circuit depth of pUNN is approximately 20.
pUNN frequently achieves or approaches the chemical accuracy threshold of 1.6 mHartree, as indicated by the shaded area on the graph.
By comparing pUNN and pUCCD, we find that, the mean absolute error (MAE) decreases from 51.9 mHartree for pUCCD to 0.6 mHartree for pUNN.
This corresponds to a reduction in error by two orders of magnitude. 
The MAE of pUNN is comparable to the MAE of UCCSD, which is 1.9 mHartree.

\begin{figure}[tbh]
\includegraphics[width=\linewidth]{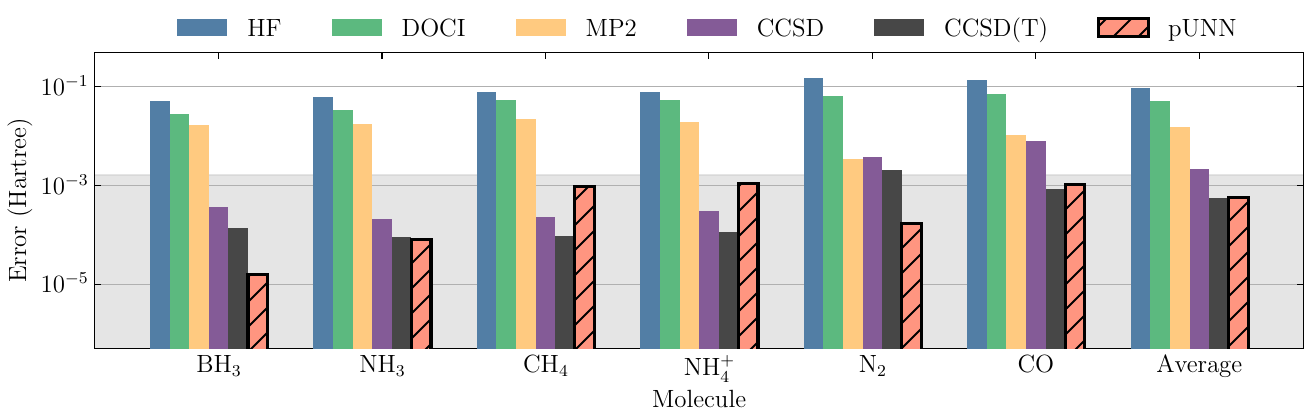}
\caption{\label{fig:error-classical} 
Compare the accuracy of pUNN with other classical computational chemistry methods.
The $1s$ orbitals are frozen and the reference energy is FCI.
The shaded area indicates the chemical accuracy.
}
\end{figure}

In Fig.~\ref{fig:error-classical} we compare the error of pUNN with several classical computational methods.
The doubly occupied configuration interaction (DOCI) method is the classical counterpart of the pUCCD method
since it also assumes electron pairing. 
Based on the results in Fig.~\ref{fig:error-quantum} we can expect  DOCI will perform poorly, which is confirmed by the data in Fig.~\ref{fig:error-classical}.
The second order Møller–Plesset perturbation theory (MP2) improves over DOCI, particularly for diatomic molecules.
This suggests that including the configurations with singly occupied orbitals is crucial for accurately describing the molecular wavefunction.
The coupled-cluster methods, CCSD and its perturbative extension CCSD(T), are considered some of the most accurate techniques in quantum chemistry. 
Both CCSD and CCSD(T) demonstrate high accuracy, with CCSD(T) achieving chemical accuracy for most of the molecules studied.
When comparing pUNN to these classical methods, we find that pUNN achieves accuracy comparable to that of CCSD(T), indicating that pUNN is a high-accuracy method for quantum chemistry calculations.

\begin{figure}[tbh]
\includegraphics[width=\linewidth]{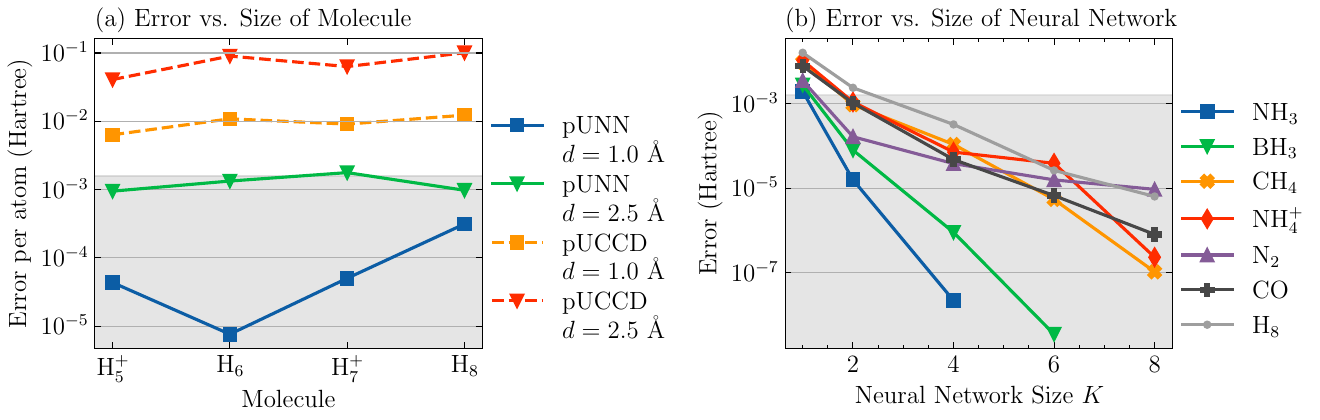}
\caption{\label{fig:error-scaling} 
Factors for the accuracy of the pUNN method.
(a) The error of pUNN versus the size of the molecule under study.
(b) The error of pUNN versus the size of the neural network $K$.
The number of hidden neurons in the neural network is $2KN$ where $N$ is the number of molecular orbitals.
The shaded area indicates the chemical accuracy.
}
\end{figure}

We next investigate the factors that determine the accuracy of the pUNN method.
Fig.~\ref{fig:error-scaling}(a) compares the accuracy of pUCCD and pUNN methods against the size of hydrogen chain molecules (\ch{H5+}, \ch{H6}, \ch{H7+}, and \ch{H8}) for two different bond lengths ($d$ = 1.0 Å and $d$ = 2.5 Å). The results clearly demonstrate that pUNN consistently outperforms standard pUCCD, achieving lower error across all molecule sizes and bond lengths. 
Notably, pUNN maintains high accuracy even as the molecule size increases especially for the longer bond length of 2.5 Å.
When $d$= 1.0 Å, the errors of pUNN seem to fluctuate when the system size varies. 
However, the magnitude of the fluctuation, in the order of $10^{-4}$ Hartree, is well below the chemical accuracy threshold and thus insignificant.
Fig.~\ref{fig:error-scaling}(b) showcases the impact of neural network size on the error of pUNN for various molecules. 
The $x$-axis represents the neural network size $K$, and the number of hidden neurons is $2KN$
where $N$ is the number of molecular orbitals. 
The atomic distance in \ch{H8} is $d$ = 1.0 Å.
As the network size increases from 2 to 8, there's a clear trend of a logarithmic decreasing error for all molecules. 
Most molecules achieve chemical accuracy (indicated by the shaded area) with larger neural networks, with \ch{NH3} and \ch{BH3} showing particularly significant improvements in accuracy as the network size grows.
Although molecules studied here are still much smaller than those encountered in practical chemistry problems,
the promising scaling shown in Fig.~\ref{fig:error-scaling}(a) suggests that pUNN has the potential to accurately learn the wavefunction of complex chemical systems.

\begin{figure}[tbh]
\includegraphics[width=0.9\linewidth]{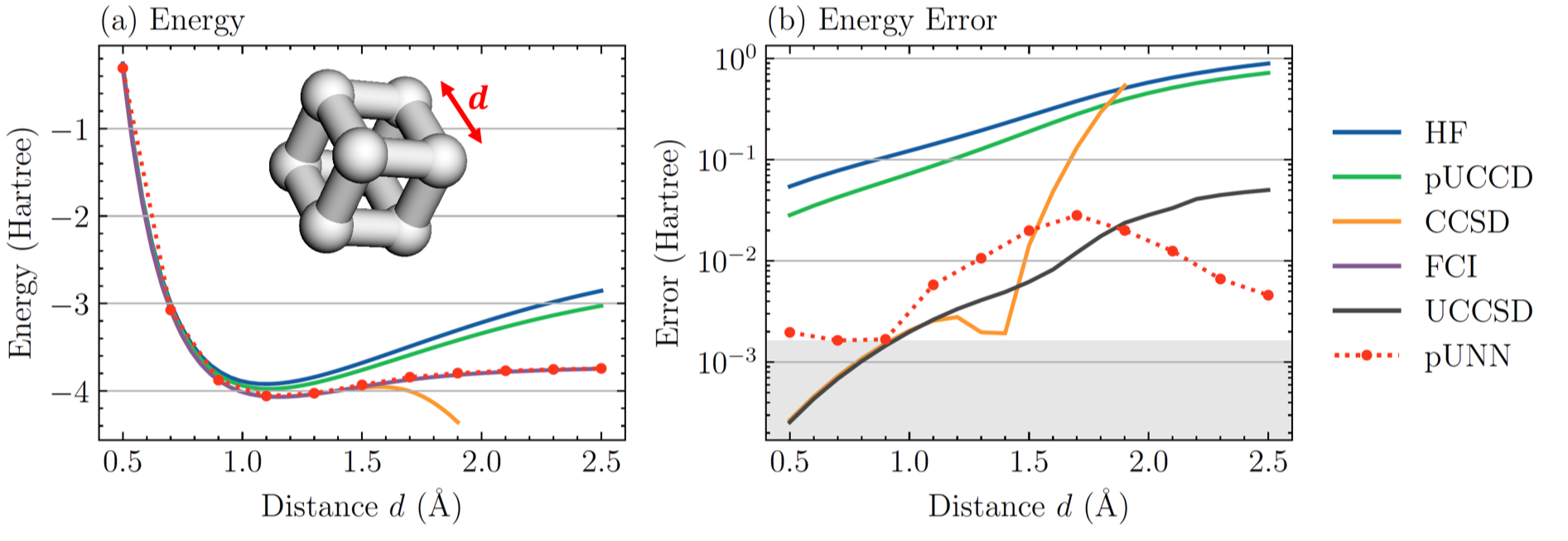}
\caption{\label{fig:h8}
Benchmarking pUNN based on the potential energy profile of cubic \ch{H8}.
(a) The potential energy profile of cubic \ch{H8} by different computational methods.
(b) The error compared with the exact solution versus the H-H distance in the \ch{H8} cube.
}
\end{figure}

We finally test the accuracy of pUNN based on cubic \ch{H8} molecule at different H-H distance $d$.
The system is particularly challenging due to the strong correlation as $d$ increases.
In Fig.~\ref{fig:h8}(a) we show the potential energy profile computed by both pUNN.
As expected, pUNN shows much higher accuracy than other methods.
From $d=0.5$ Å to 2.5 Å pUNN coincides well with the FCI solution.
For reference, we also include the CCSD method, which shows high accuracy at intermediate $d$.
However, due to its single-reference and non-variational nature, the error of CCSD quickly increases as $d$ becomes larger than 1.5 Å and it fails to reach convergence for larger $d$.
CCSD(T) is not expected to improve CCSD when it fails because CCSD(T) relies on good CCSD wavefunction to account for perturbative triple excitation.
\wtli{
Thus, although the 16 qubit system represents a relatively small variational space compared to challenging strongly correlated systems~\cite{li2019electronic, larsson2022chromium},
it is sufficient to reveal the limitations of methods like CCSD, which fail in strongly correlated regimes, while pUNN maintains relatively high accuracy.
}
In Fig.~\ref{fig:h8}(b) we depicted the error of the methods in logarithmic scale.
All methods except pUNN show an increase in error as $d$ increases.
The maximum error for pUNN appears at $d=1.7$ Å and the magnitude of the error is $10^{-2}$ Hartree.
The relatively large error highlights the complexity of the cubic \ch{H8} molecule.
\wtli{We anticipate that integrating alternative quantum circuits into our pUNN framework, such as those based on valence bond theory~\cite{hu2025unitary}, could enhance accuracy in strong correlation}.
The UCCSD method is also included in Fig.~\ref{fig:h8}(b). 
While UCCSD shows high accuracy at smaller $d$, it suffers from significant error at the large $d$ limit, similar to CCSD.
\wtli{
We perform additional benchmarks for strongly correlated systems based on the potential energy profile of \ch{N2} and \ch{CH4} and the trend is similar to Fig.~\ref{fig:h8}(b).
The results are included in the Supplementary Information.
}

\wtli{
In Table~\ref{tab:params}, we present a breakdown of parameters for hydrogen systems studied in Fig.~\ref{fig:error-scaling} and in Fig.~\ref{fig:h8}, comparing pUNN with FCI. 
For pUNN, the pUCCD circuit has $\mathcal{O}(N^2)$ parameters, while the NN has $\mathcal{O}(K^2N^2L)$ parameters, with $K=2$ and $L=N-3$. 
From Table~\ref{tab:params}, pUNN’s total parameters grow polynomially with $N$, while FCI’s determinant space grows exponentially. 
For \ch{H8}, pUNN uses fewer parameters than FCI,
and achieves high accuracy across both weak and strong correlation,
as shown in Fig.~\ref{fig:error-scaling}(a) and Fig.~\ref{fig:h8}. 
As our main contribution is the novel and unique quantum-neural hybrid framework, 
our choice of a dense MLP for the neural network is a proof-of-concept.
More efficient architectures, such as restricted Boltzmann machines or graph neural networks, could further optimize pUNN~\cite{carleo2017solving, pescia2024message}, 
}
\begin{table}[ht]
\caption{\wtli{Parameter Counts for pUNN and FCI for Hydrogen Systems}}
\label{tab:params}
\begin{center}
\begin{tabular}{cccc}
\hline
System & FCI Determinants & NN Parameters & pUCCD Parameters \\
\hline
\hline
H$_5^+$ & 100 & 661 & 6 \\
H$_6$ & 400 & 1537 & 9 \\
H$_7^+$ & 1225 & 2885 & 12 \\
H$_8$ & 4900 & 4801 & 16 \\
H$_{2n}$ & $\left[(2n)!/(n!)^2\right]^2$ & $128n^3-208n^2-16n+1$ & $n^2$ \\
\hline
\end{tabular}
\end{center}
\end{table}

\subsection{Experiments on a Superconducting Quantum Computer}
\begin{figure}[tbh]
\includegraphics[width=\linewidth]{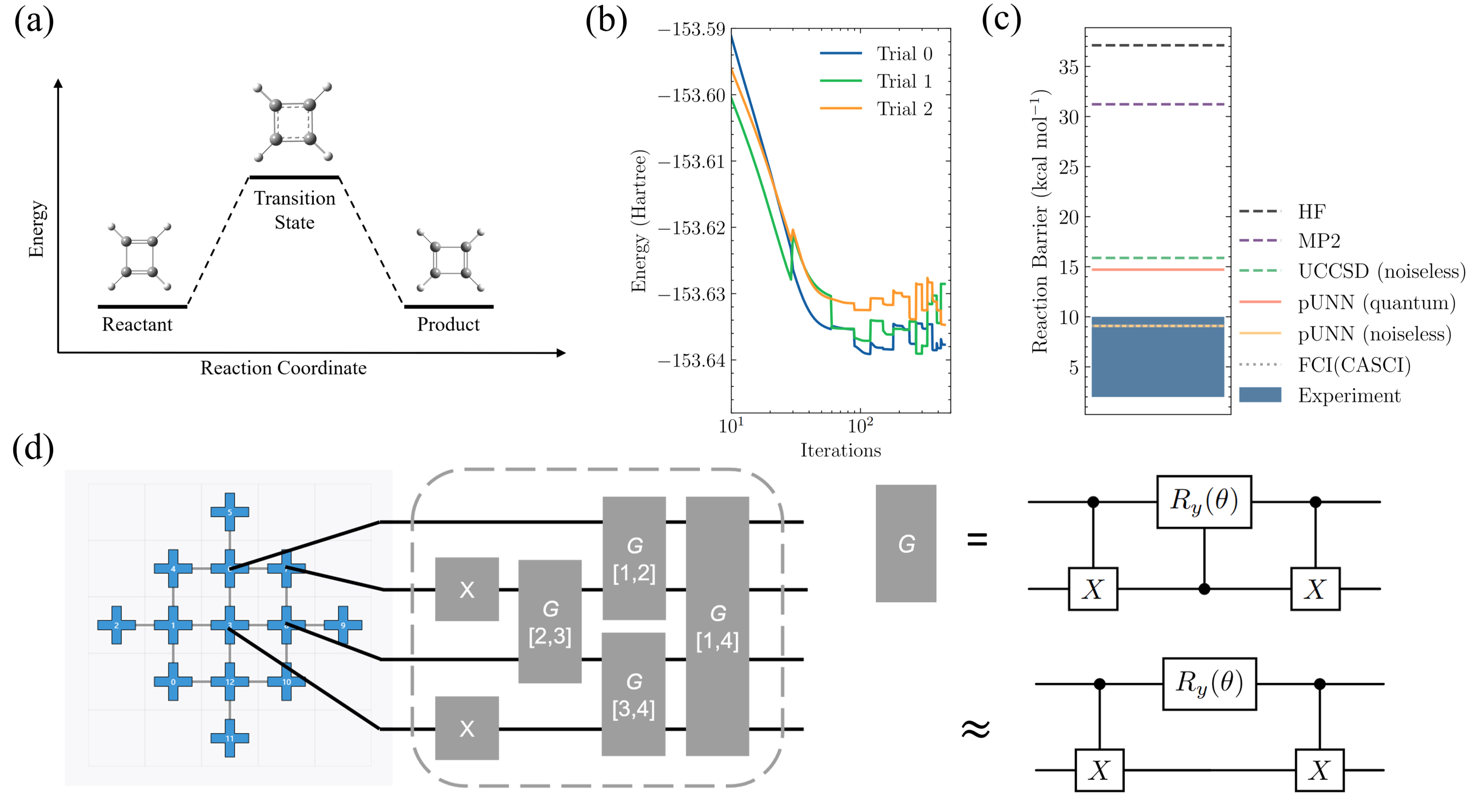}
\caption{\label{fig:qc}
Experiments on a superconducting quantum computer.
(a) The isomerization reaction of cyclobutadiene, with the transition state energy calculated using pUNN on a superconducting quantum computer.
(b) The estimated energy during the optimization process. Results by three independent random initializations of the neural network are shown.
(c) The computed reaction barrier from pUNN, compared with results from several other computational methods. 
``Experiment'' means the reaction barrier calculated by experimentally observed chemical reaction rate.
(d) The 13-qubit superconducting quantum chip and the quantum circuit used for the calculation.
}
\end{figure}

To evaluate the performance of pUNN in a real quantum computing scenario, we conduct experiments on a superconducting quantum computer. 
We choose the isomerization reaction of cyclobutadiene as our model system, as shown in Fig.~\ref{fig:qc}(a). 
The transition state of this system is particularly challenging due to strong correlations arising from degeneracy~\cite{lyakh2012multireference, hermann2020deep}. 
In this reaction, the reactant and product are identical molecules, with a 90-degree rotation between them. The electronic structures of the reactant and the product are considerably simpler than that of the transition state. Therefore, in the following analysis, we focus on the transition state,
and calculate the reaction barrier by subtracting the exact energy of the reactant and product from the energy of the transition state.

We employ the cc-pVDZ basis set~\cite{Dunning89} for HF calculation and select the four frontier orbitals as the active space. Using the paired ansatz, the active space is represented by a 4-qubit quantum circuit, with four parameters corresponding to four double excitations.
The superconducting quantum chip used in this work consists of 13 qubits. Since the Givens-Swap gate is not a native gate on this chip, we carefully select 4 qubits from the 13-qubit system, which follows a ring topology, as shown in Fig.~\ref{fig:qc}(d). This selection allows us to implement all four excitation operators using only Givens rotation gates, eliminating the need for the more expensive swap gates, which would otherwise require 3 CNOT gates.
The Givens rotation gates should be further compiled into 4 native CNOT gates, along with several single-qubit gates. To reduce circuit depth, we introduce an approximation that breaks the symmetry and removes the control qubit of the controlled $R_y$ gate~\cite{magoulas2023linear}. 
The resulting circuit does not conserve the total particle number anymore but the overall error could be smaller than the gate error by 8 additional CNOT gates, especially when some of the rotation gates have small rotation angles.
Each Givens rotation gate is thus compiled into 2 CNOT gates, resulting in a total of 8 CNOT gates in the circuit.
Standard readout error mitigation based on a direct product calibration matrix is applied to enhance the precision.

We obtain the circuit parameters by optimizing the pUCCD Hamiltonian on this chip using the SOAP optimizer~\cite{li2024efficient}, 
which is an efficient optimizer tailored for parameter optimization on quantum circuits.
Next, we train a neural network based on the sampling output from the optimized quantum circuit. 
In Fig.~\ref{fig:qc}(b), we report the energy estimates during the optimization process. 
Sampling from the quantum circuit occurs every 30 steps, with the macro iteration performed 15 times, for a total of 450 iterations. 
The number of iterations is determined by trial classical simulation, which ensures convergence.
For each quantum circuit, we perform 1024 shots of measurement for each Pauli string.
The optimization is repeated with three different neural network initializations and the lowest energy is employed for reaction barrier calculation.

As shown in Fig.~\ref{fig:qc}(c), the reaction barrier predicted by pUNN on the quantum circuit is approximately 16 $\textrm{kcal mol}^{-1}$. While this value is still higher than the experimentally reported range of $2 \sim 10 \ \textrm{kcal mol}^{-1}$~\cite{whitman1982limits}, it represents a notable improvement over the HF and MP2 energies, and is comparable to the noiseless UCCSD prediction.
When using a noiseless pUNN model, obtained via a statevector simulator, the predicted reaction barrier is around 9 $\textrm{kcal mol}^{-1}$, which aligns well with the FCI results and experimental observations. 
This highlights the importance of addressing errors introduced by quantum circuit gates and measurement uncertainties. 
In particular, the neural network parameters with quantum computers are different from the neural network parameters with noiseless simulation. We conjecture that pUNN(quantum) predicts a higher energy because the neural network parameters are stuck in a local minimum.
To improve the performance of pUNN in the presence of these errors, advanced optimizers, such as KFAC~\cite{martens2015optimizing, pfau2020ab}, could be considered. 
Yet adaption of the KFAC optimizer will likely be necessary due to the unique algorithmic structure of pUNN.


\begin{figure}[tbh]
\includegraphics[width=0.6\linewidth]{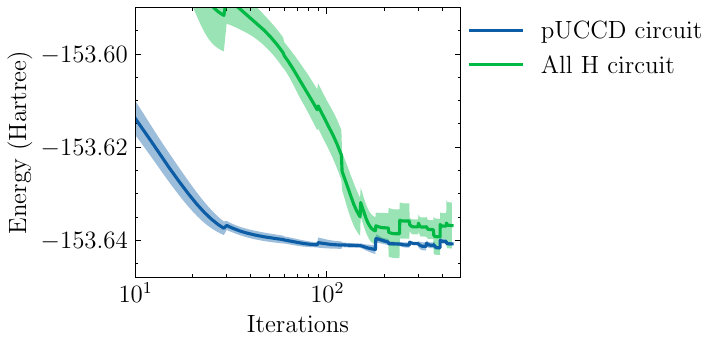}
\caption{\label{fig:qc_advantage}
Energy estimates during neural network training with different quantum circuits.
This figure illustrates the effect of quantum circuits on energy estimation within the pUNN framework: (1) the pUCCD circuit, which is the circuit used throughout this paper, and (2) the Hadamard superposition circuit, where Hadamard gates are applied to all qubits, creating a superposition of all possible states. The standard deviation across five different neural network initializations is shown as the shaded area.
}
\end{figure}

Next, we investigate the advantage of incorporating quantum computing into the pUNN framework. Since neural networks are widely known for their effectiveness across a variety of tasks including representing molecular wavefunction, it is important to assess whether a quantum circuit is truly necessary for this framework. To explore this, we replace the pUCCD circuit in pUNN with a Hadamard superposition circuit, where Hadamard gates are applied to all qubits. The Hadamard superposition circuit can be easily emulated on classical computers and can be considered as a ``dummy'' sample generator when used to compute the energy with the neural network. To isolate the impact of quantum gate noise, we perform the comparison using a shot-based classical emulator which is free of gate noise.
We use the transition state of the cyclobutadiene isomerization reaction as our model system.
As shown in Fig.~\ref{fig:qc_advantage}, replacing the pUCCD circuit with a Hadamard superposition leads to a noticeable decrease in accuracy, along with a significant increase in energy variance. In fact, for large molecules, a Hadamard superposition circuit greatly reduces the probability of sampling the dominant configuration, making it less effective for energy estimation. In contrast, the pUCCD circuit provides a suitable starting point for further refinement through neural network training, demonstrating the advantage of quantum computing in this context.

\section{Conclusion and Outlook}
The pUNN framework combines an efficient quantum circuit with the expressive power of a neural network to accurately and robustly compute molecular energies. 
Through a carefully designed algorithmic structure—including the pUCCD circuit, entanglement circuit, perturbation circuit, and neural network augmentation—the method achieves high accuracy with low quantum resource requirements, utilizing only $N$ qubits instead of the $2N$ qubits typically required by comparable methods. 
The incorporation of a neural network allows the framework to mitigate errors effectively, making it robust to gate noise and capable of delivering consistent accuracy on noisy quantum hardware.
The design also ensures manageable measurement overhead for the interaction between the quantum circuit and the neural network.

Extensive numerical benchmarks demonstrate that pUNN achieves accuracy comparable to advanced methods like UCCSD, while being more resource-efficient and scalable to larger molecular systems. 
Experimental validation on a superconducting quantum computer demonstrates the practicality of this approach. 
With a 4-qubit quantum circuit, pUNN successfully computes the transition state energy of cyclobutadiene isomerization, yielding energy estimates with accuracy comparable to noiseless UCCSD. 
Based on this model reaction, we also demonstrate that the quantum circuit plays an indispensable role in the hybrid framework, as replacing it with a neural network alone leads to a higher error and, crucially, a significantly larger variance in energy estimation.
\wtli{This observation serves as an evidence for the advantage of this hybrid design than pure classical neural networks,
where the quantum circuit reduces the representational burden on the neural network.
Thus, we expect that pUNN is able to demonstrate quantum advantage as we tackle larger systems where classical simulation of the pUCCD circuit becomes intractable.}

While this work focuses on closed-shell systems, the pUNN framework can be directly extended to open-shell systems by modifying the particle number conservation mask in the neural network. However, since the pUCCD quantum circuit may not accurately approximate open-shell wavefunctions, further adaptations will likely be necessary to maintain accuracy for open-shell systems.
Future work could enhance the neural network architecture by incorporating more sophisticated neural layers with physical insights. Additionally, pretraining the neural network on a diverse set of molecules offers a possible avenue for creating a generalizable model that can be fine-tuned for specific systems.

\section*{Appendix A: The electronic structure problem and the \NoCaseChange{p}UCCD ansatz}
In this work,
we are interested in the second-quantized \textit{ab initio} electronic structure Hamiltonian
\begin{equation}
\label{eq:ham-abinit}
    \hat H = \sum_{pq}h_{pq} \hat a^\dagger_p \hat a_q +\frac{1}{2}\sum_{pqrs}h_{pqrs}\hat a^\dagger_p \hat a^\dagger_q \hat a_r \hat a_s + E_{\rm{nuc}}, \
\end{equation}
where $h_{pq}$ and $h_{pqrs} = [ps|qr]$ are one-electron and two-electron integrals, and $\hat a^\dagger_p, \hat a_p$ are fermionic creation and annihilation operators, respectively, acting on the $p$-th spin-orbital.

In order to compute the expectation of Eq.~\eqref{eq:ham-abinit} on a programmable quantum computer, the symmetry of the creation and annihilation operators has to be taken care of.
Creation and annihilation operators for fermions obey the anticommutation relations
\begin{equation}
 \begin{aligned}
    \{\hat a_i, \hat a^\dagger_j\} & = \delta_{ij} \\
     \{\hat a^\dagger_i, \hat a^\dagger_j\} & =  \{\hat a_i, \hat a_j\} = 0
\end{aligned}
\end{equation}
On the other hand, the qubit creation operator $\hat c^\dagger = \frac{1}{2}(X - iY)$ and annihilation operator $\hat c =\frac{1}{2} (X+iY)$ obey the commutation relations
\begin{equation}
    \{\hat c_i, \hat c^\dagger_j\} = \delta_{ij} , \quad [\hat c^\dagger_i, \hat c^\dagger_j] =  [\hat c_i, \hat c_j] = 0.
\end{equation}
In this work, when necessary, we employ the Jordan-Wigner transformation to map fermionic ladder operators into qubit operators

In general, UCC types of ansatz can be written as
\begin{equation}
\label{eq:ucc2}
    \ket{\Psi(\theta)} = \prod e^{\theta_k \hat G_k} \ket{\phi}. \
\end{equation}
Here, $\ket{\phi}$ is the Hartree--Fock state.
For the UCCSD method, $\hat G_k$ has the form
\begin{equation}
\label{eq:g-single-double}
    \hat G_k = \begin{cases}
        \hat a^\dagger_p \hat a_q - \text{h.c.} , \\
        \hat a^\dagger_p \hat a^\dagger_q \hat a_r \hat a_s - \text{h.c.}
        \end{cases}
\end{equation}

pUCCD is an efficient ansatz requiring only $\order{N}$ circuit depth and half as many qubits as other UCC ansatze~\cite{henderson2015pair, elfving2021simulating}.
pUCCD allows only paired double excitations, 
which enables one qubit to represent one spatial orbital instead of one spin orbital,
and removes the need to perform the fermion-qubit mapping.
The subspace in which all states have paired configuration is called the seniority-zero subspace.
In this subspace, there are $\order{N^2}$ double excitations, which can be executed on a quantum computer efficiently using a compact circuit.
The circuit is composed of a linear depth of Givens-SWAP gates, assuming linear qubit connectivity~\cite{elfving2021simulating}.
In the seniority-zero subspace, the Hamiltonian also takes a simpler form, with only $N^2$ terms:
\begin{equation}
\label{eq:ham-puccd}
    \hat H = \sum_p h_p \hat c^\dagger_p \hat c_p + \sum_{pq} v_{pq} \hat c^\dagger_p \hat c_q + \sum_{p \neq q} w_{pq} \hat c^\dagger_p \hat c_p \hat c^\dagger_q \hat c_q + E_{\rm{nuc}} \ ,
\end{equation}
where $h_p = 2h_{pp}$, $v_{pq} = (pq|pq)$ and $\omega_{pq} = 2 (pp|qq) - (pq|pq)$.
Here $p$ and $q$ are indices for spatial orbitals.
If we use $\hat n_p = \hat c^\dagger_p \hat c_p = \frac{1-Z}{2}$ to denote occupation number operator,
Eq.~\eqref{eq:ham-puccd} can be converted to a sum of Pauli string where the maximum length of Pauli string is 2.
Meanwhile, the first and the third term in Eq.~\eqref{eq:ham-puccd} have only $Z$ terms and the second term will contribute to $XX$ and $YY$ terms.
Thus, the expectation of Eq.~\eqref{eq:ham-puccd} can be measured in 3 different bases, regardless the number of qubit involved.

\section{Appendix B: The measurement protocol for \NoCaseChange{p}UNN}
\label{app:measure}
To begin with, we describe the measurement protocol when a single quantum circuit is integrated with a neural network, following the reference~\cite{zhang2022variational}.
Then, we move on to our measurement method that enables efficient measurement of two separate circuits in the pUNN algorithm, defined in Eq.~\eqref{eq:two-circuit-measure}.
In the following, for clarity, we omit the prime symbol for both $\hat H'$ and $\hat N'$,
since $\hat H'$ is Pauli string similar to $\hat H$, and $\hat N'$ follows the definition of $\hat N$ in Eq.~\eqref{eq:def-nn}.

\subsection{A single quantum circuit}
\label{sec:measure-1}

For a single circuit $\ket{\psi} = \sum_k a_k\ket{k}$, where $\ket{k}$ is the computational basis,
$\hat N$ is written as
\begin{equation}
    \hat N = \sum_k b_k \ket{k}\bra{k} \ .
\end{equation}
We then focus on deriving an appropriate form of $\hat N \hat H \hat N$.
We assume that both $\ket{\psi}$ and $\hat N$ are real-valued. 
We first derive the measurement protocol for the norm of $\ket{\Psi}=\hat N \ket{\psi}$, given by
\begin{equation}
    \braket{\Psi|\Psi} = \braket{\psi|\hat N^\dagger \hat N|\psi} \ ,
\end{equation}
where
\begin{equation}
\label{eq:nn}
    \hat N^\dagger \hat N = \sum_k b_k^2 \ket{k}\bra{k} \ .
\end{equation}
Clearly, the eigenvectors of $\hat N^\dagger \hat N$ are $\ket{k}$ and their eigenvalues are $b^2_k$.
To compute the norm, we sample bitstrings from $\ket{\psi}$ and multiply the probability of $k$ by $b^2_k$.
For efficient sampling, $b_k$ should not be too large or small. In other words, $a_k$ must provide a good first-order approximation to the ground state. The same is also true for our measurement protocol for 2 circuits and it highlights the role of the quantum computer in this framework.

\newcommand{\ka}{{\tilde{k}}}
\newcommand{\ja}{{\tilde{j}}}

Next, we consider the measurement of a Pauli string $\hat H$.
The main focus is to derive $\hat N^\dagger \hat H \hat N$.
In general, a Pauli string $\hat H$ can be written as
\begin{equation}
    \hat H = \sum_k S_\ka \ket{\ka}\bra{k} \ ,
\end{equation}
where the summation is over $k$ rather than $k$ and $\ka$.
In other words, applying the Pauli string $\hat H$ on $\ket{k}$ will produce only one bitstring $\ket{\ka}$
up to a phse $S_{\ka}$
\begin{equation}
\label{eq:ham-one-bitstring}
    \hat H \ket{k} = S_{\ka} \ket{\ka} \ .
\end{equation}
Since $\hat H^2 = I$, we also have $\hat H \ket{\ka} = S_{k}\ket{k}$ and $S_{\ka} S_k = 1$.

Let's first consider the case where $\hat H$ has only $Z$ operators, i.e. $\ket{k}=\ket{\ka}$.
In this case, the term to measure is 
\begin{equation}
\label{eq:nzn}
    \hat N^\dagger \hat H \hat N = \sum_k b_k^2 S_k \ket{k}\bra{k} \ .
\end{equation}
Eq.~\eqref{eq:nzn} is similar to the expression for $\hat N^\dagger \hat N$ in Eq.~\eqref{eq:nn}.
As a result, the measurement protocol when $\hat H$ only involves $Z$ operators is very similar to the procedure for measuring the norm of the state.

Now, consider the general case where $\hat H$ includes at least one $X$ or $Y$ operator,
where it is ensured that $\ket{k} \neq \ket{\ka}$. 
In this case, we can rewrite $\hat H$ as 
\begin{equation}
    \hat H  = \sum_{k \in \Omega} \left (S_k\ket{k}\bra{\ka} + S_\ka\ket{\ka}\bra{k} \right ) \ ,
\end{equation}
where $\Omega=\{k|\textrm{bin}(k)<\textrm{bin}(\ka)\}$ and $\textrm{bin}(k)$ refers to the corresbonding binary integer of $k$.

The Hamiltonian transformed by $\hat N$ is given by
\begin{equation}
\label{eq:nhn-1q}
    \hat N^\dagger \hat H \hat N = \sum_{k \in \Omega} b_kb_\ka \hat H_k \ ,
\end{equation}
where
\begin{equation}
        \hat H_k  = S_k\ket{k}\bra{\ka} + S_\ka\ket{\ka}\bra{k} \ .
\end{equation}
To measure $ \hat N^\dagger \hat H \hat N$, we need to derive the eigenvectors of $\hat H_k$.
$\hat H_k$ is defined by two basis $\ket{k}$ and $\ket{\ka}$ and therefore
$\hat H_k$ has two eigenvectors with eigenvalues +1 and -1.
Denote the two eigenvectors as $\ket{k^+}$ and $\ket{k^-}$, we can then write $\hat H_k$ as
\begin{equation}
    \hat H_k = \ket{k^+}\bra{k^+} - \ket{k^-}\bra{k^-} \ .
\end{equation}
In the computational basis, $\ket{k^+}$ and $\ket{k^-}$ are written as
\begin{equation}
\label{eq:1h-eigen}
\begin{aligned}
    \sqrt{2}\ket{k^+} = S_\ka \ket{\ka} + \ket{k} = (\hat H_k + 1) \ket{k} \ , \\
    \sqrt{2}\ket{k^-} = S_\ka \ket{\ka} - \ket{k} = (\hat H_k - 1) \ket{k} \ .
\end{aligned}
\end{equation}
These eigenvectors have eigenvalues $+1$ and $-1$, respectively.
The neural network transformed Hamiltonian is then
\begin{equation}
\label{eq:nn-ham}
    \hat N^\dagger \hat H \hat N  =\sum_{k \in \Omega} b_k b_\ka \left (\ket{k^+}\bra{k^+} - \ket{k^-}\bra{k^-} \right )
     \ .
\end{equation}

To perform the measurement in the $\ket{k^+}$ and $\ket{k^-}$ bases, 
we append a unitary measurement circuit $V$ to the original quantum circuit $\ket{\psi}$.
$V$ satisfies
\begin{equation}
\label{eq:def-v}
\begin{aligned}
    V^\dagger \ket{k} =  \ket{k^+} \ , \\
    V^\dagger \ket{\ka} = \ket{k^-} \ ,
\end{aligned}
\end{equation}
for any $k \in \Omega$. The unitary property can be proven by considering $\braket{k'|VV^\dagger|k}$ or by noting that $V$ is a permutation between two sets of orthonormal basis states.
The construction of the transformation circuit $\hat V$ is a standard procedure in quantum computation,
because $\hat V$ is a circuit that diagonalizes the Pauli string $\hat H$.
If the number of $X$ and $Y$ operators in $\hat H$ is $m$, then the number of two-qubit gates in $\hat V$ is $m-1$.

To summarize, the quantum circuit used for the measurement is $\hat V\ket{\psi}$, and the term to measure is
\begin{equation}
    \hat V \hat N^\dagger \hat H \hat N \hat V^\dagger =  \sum_{k \in \Omega} b_k b_\ka \left (\ket{k}\bra{k} - \ket{\ka}\bra{\ka} \right ) \ .
\end{equation}
The expectation value of this term is readily accessible from the quantum circuit $\hat V\ket{\psi}$ by performing a projection measurement in the computational basis.

\subsection{Two separate quantum circuits}
If we take the two separate quantum circuit $\ket{\psi \otimes \phi}$ as a whole,
the measurement protocol developed in Sec.~\ref{sec:measure-1} can be applied to
measure the expectation when a neural network is integrated with $\ket{\psi \otimes \phi}$.
However, in this case, the unitary transformation for measurement $V$ will generally entangle the two originally unentangled quantum circuits.
This results in a quantum circuit of $2N$ qubits.
If we wish to avoid this entanglement and measure the expectation using two separate quantum circuits, a special measurement procedure is needed. This procedure will be described in the following.

The total wavefunction is expressed as:
\begin{equation}
    \ket{\Psi} = \hat N \ket{\psi \otimes \phi} \ .
\end{equation}
In the pUNN framework, $\ket{\psi}$ is the pUCCD quantum circuit, and $\ket{\phi}$ is the perturbation circuit to be simulated classically.
However, the procedure outlined below is general and can be readily applied to other cases involving uncorrelated circuits.

Consider Hamiltonian in the form:
\begin{equation}
    \hat H = \hat H_{\psi} \otimes \hat H_{\phi} \ ,
\end{equation}
where $\hat H_{\psi}$ and $\hat H_{\phi}$ are Pauli strings for the two separate circuits.
If either of  $\hat H_{\psi}$ and $\hat H_{\phi}$ does not contain $X$ or $Y$, 
the measurement procedure simplifies to the standard approach described in Sec.~\ref{sec:measure-1}.
Therefore, we will focus on the general case where
both $\hat H_{\psi}$ and $\hat H_{\phi}$ contain $X$ or $Y$.
Similar to Eq.~\eqref{eq:ham-one-bitstring},
$\hat H_{\psi}$ and  $\hat H_{\phi}$ satisfy the following relations:
\begin{equation}
\label{eq:h-two}
\begin{aligned}
    \hat H_\psi \ket{k} & = S_{\ka} \ket{\ka} \ , \\
    \hat H_\phi \ket{j} & = S_{\ja} \ket{\ja} \ .
\end{aligned}
\end{equation}
Here, $\hat H_{\psi}$ and  $\hat H_{\phi}$ act independently on the circuit $\ket{\psi}$ and $\ket{\phi}$,
transforming the states $\ket{k}$ and $\ket{j}$ into $\ket{\ka}$ and $\ket{\ja}$, with corresponding signs $S_{\ka}$ and $S_{\ja}$.

The eigenvectors of $\hat H$ are given by
\begin{equation}
\begin{aligned}
    2\ket{k^\pm} \ket{j^\pm} = \left ( S_{\ka}\ket{\ka} \pm \ket{k} \right )
     \left ( S_{\ja}\ket{\ja} \pm \ket{j} \right ) \ ,
\end{aligned}
\end{equation} 
where we again require $k \in \Omega_\psi$ and $j \in \Omega_\phi$ to avoid double-counting.
In the following, we use $k, j \in \Omega$ as a short-hand notation for the condition.

The Hamiltonian in the computational basis is
\begin{equation}
\begin{aligned}
    \hat H  & = \sum_{k, j \in \Omega} \left (S_{k}\ket{k}\bra{\ka} + S_{\ka}\ket{\ka}\bra{k} \right ) \otimes \left (S_{j}\ket{j}\bra{\ja} + S_{\ja}\ket{\ja}\bra{j} \right ) \\
    & =  \sum_{k, j \in \Omega} \left ( S_{k}S_{j}\ket{k, j}\bra{\ka, \ja} + S_{\ka}S_{\ja}\ket{\ka, \ja}\bra{k, j}
    \right ) \\
    & \quad + \sum_{k, j \in \Omega} \left ( 
    S_{k}S_{\ja}\ket{k,\ja}\bra{\ka,j} + S_{\ka}S_{j}\ket{\ka,j}\bra{k,\ja} 
    \right )  \ .
\end{aligned}
\end{equation}
After applying the transformation $\hat N$, the transformed Hamiltonian becomes
\begin{equation}
\begin{aligned}
\label{eq:nhn-2q}
    \hat N^\dagger \hat H  \hat N
    & =  \sum_{k, j \in \Omega} b_{kj} b_{\ka\ja} \left ( S_{k}S_{j}\ket{k, j}\bra{\ka, \ja} + S_{\ka}S_{\ja}\ket{\ka, \ja}\bra{k, j}
    \right ) \\
    & \quad + \sum_{k, j \in \Omega} b_{k\ja} b_{\ka j} \left ( 
    S_{k}S_{\ja}\ket{k,\ja}\bra{\ka,j} + S_{\ka}S_{j}\ket{\ka,j}\bra{k,\ja} 
    \right ) \ .
\end{aligned}
\end{equation}
The structure of $\hat N^\dagger \hat H  \hat N$ remains similar to $\hat H$, but the terms are now weighted by the neural network coefficients $b_{kj}$.
Eq.~\eqref{eq:nhn-2q} is more complex than Eq.~\eqref{eq:nhn-1q} since each term can not be readily factored into the direct product of operators acting on the two separate circuits. 
Consequently, finding a measurement circuit that diagonalizes Eq.~\eqref{eq:nhn-2q} without introducing entanglement between the two circuits is not straightforward.

To proceed, it is instructive to consider a 2-qubit system and with the Hamiltonian $\hat H = XX$ as an example.
In this case, we can express the neural network transformed Hamiltonian as:
\begin{equation}
\begin{aligned}
    \hat N^\dagger \hat H \hat N & = b_{00} b_{11} (\ket{00}\bra{11} + \ket{11}\bra{00}) + b_{01} b_{10}  (\ket{01}\bra{10} + \ket{10}\bra{01}) \\
    & = \frac{1}{2} b_{00} b_{11} (XX-YY) +  \frac{1}{2}b_{01} b_{10} (XX+YY) \\
    & = \frac{1}{2} ( b_{00} b_{11} +  b_{01} b_{10})XX + \frac{1}{2}( -b_{00} b_{11} +  b_{01} b_{10})YY \ .
\end{aligned}
\end{equation}
Thus, to measure the expectation value of $XX$ in the presence of a NN, one needs to measure both $XX$ and $YY$ to avoid measurement circuit that entangles the two separate circuits.

More generally, consider a Hamiltonian $\hat J = \hat J_\psi \otimes \hat J_\phi$ such that (as in Eq.~\eqref{eq:h-two})
\begin{equation}
\begin{aligned}
    \hat J_\psi \ket{k} & = iS_{\ka} \ket{\ka} \ , \\
    \hat J_\phi \ket{j} & = iS_{\ja} \ket{\ja}  \ .
\end{aligned}
\end{equation}
$\hat J$ can be constructed by replacing an $X$ operator with $-Y$ or a $Y$ operator with $X$ in $\hat H_\psi$ and $\hat H_{\phi}$.
The eigenvectors of $\hat J$ are
\begin{equation}
\begin{aligned}
    2\ket{k^{i\pm}} \ket{j^{i\pm}} = \left (i S_{\ka}\ket{\ka} \pm \ket{k} \right )
     \left (i S_{\ja}\ket{\ja} \pm \ket{j} \right ) \ .
\end{aligned}
\end{equation} 
We define short-hand notation for the projectors
\begin{equation}
    \hat h_{k}^\pm = \ket{k^\pm} \bra{k^\pm} \,
\end{equation}
which form the diagonal bases for $\hat H$ and $\hat J$.
The first term of $\hat N^\dagger \hat H \hat N$ from Eq.~\eqref{eq:nhn-2q} is then transformed to:
\begin{equation}
\begin{aligned}
    &   S_{k} S_{j} \ket{k, j}\bra{\ka, \ja} + S_{\ka} S_{\ja}\ket{\ka, \ja}\bra{k, j}  \\
    & = \frac{1}{2}(\hat h_{k}^+ - \hat h_{k}^-)\otimes (\hat h_{j}^+ - \hat h_{j}^-) -
    \frac{1}{2}(\hat h_{k}^{i+} - \hat h_{k}^{i-})\otimes (\hat h_{j}^{i+} - \hat h_{j}^{i-}) \ .
\end{aligned}
\end{equation}
Similarly, the second term becomes
\begin{equation}
\begin{aligned}
    & S_{k}S_{\ja}\ket{k,\ja}\bra{\ka,j} + S_{\ka}S_{j}\ket{\ka,j}\bra{k,\ja}  \\
    & = \frac{1}{2}(\hat h_{k}^+ - \hat h_{k}^-)\otimes (\hat h_{j}^+ - \hat h_{j}^-) +
    \frac{1}{2}(\hat h_{k}^{i+} - \hat h_{k}^{i-})\otimes (\hat h_{j}^{i+} - \hat h_{j}^{i-}) \ .
\end{aligned}
\end{equation}
The overall expression for $\hat N^\dagger \hat H \hat N$ is
\begin{equation}
\begin{aligned}
\label{eq:two-circuit-measure-final}
    \hat N^\dagger \hat H \hat N &= \sum_{k, j \in \Omega} \frac{1}{2} \left ( b_{kj} b_{\ka\ja} + b_{k\ja} b_{\ka j} \right ) 
    (\hat h_{k}^+ - \hat h_{k}^-)\otimes (\hat h_{j}^+ - \hat h_{j}^-) \\
    & \quad + \sum_{k, j \in \Omega} \frac{1}{2} \left ( - b_{kj} b_{\ka\ja} + b_{k \ja} b_{\ka j} \right ) (\hat h_{k}^{i+} - \hat h_{k}^{i-})\otimes (\hat h_{j}^{i+} - \hat h_{j}^{i-})
\end{aligned}
\end{equation}
One may verify the equation by setting $b=1$ and $\hat N$ becomes $\hat I$. 
In this case, the second term vanishes and the first term reduces to the original Hamiltonian $\hat H$.
In Eq.~\eqref{eq:two-circuit-measure-final}, the operators are factored into the direct product of operators acting on the two separate circuits.
Consequently, they can be diagonalized to the computational basis separately following the approach discussed in Sec.~\ref{sec:measure-1}.

Thus, in order to measure the expectation in Eq.~\eqref{eq:two-circuit-measure},
one has to sample bitstrings from both $\hat H$ and $\hat J$ and calculate the expectation following Eq.~\eqref{eq:two-circuit-measure-final} accordingly.
In the framework of pUNN, we first sample bitstrings that correspond to $\hat h_{k}^\pm$ and  $\hat h_{k}^{i\pm}$ on quantum computers,
and then sample bitstrings that correspond to $\hat h_{j}^\pm$ and  $\hat h_{j}^{i\pm}$ on classical simulators.
Then we query the neural network $\mathcal{B}$ for $b_{k, j}$, and finally calculate the expectation based on the sampling statistics and the output from the neural network.

\section*{Author Contributions}
Weitang Li: Conceptualization, data curation, investigation, methodology, validation, visualization, writing – original draft, supervision, funding acquisition;

Shi-Xin Zhang: Conceptualization, formal analysis, software, writing – review \& editing;

Zirui Sheng: Investigation, writing – review \& editing;

Cunxi Gong: Investigation, visualization, writing – review \& editing;

Jianpeng Chen: Investigation, writing – review \& editing;

Zhigang Shuai: Conceptualization, investigation, resources, writing – review \& editing, supervision, project administration, funding acquisition;

\section*{Competing Interest}
The authors declare no competing interests.

\section*{Code Availability}
The code for this study is available from the TenCirChem-NG package hosted on GitHub
\url{https://github.com/tensorcircuit/TenCirChem-NG} and \wtliwtli{from Zenodo\cite{tcczenodo}}.

\section*{Data Availability}
The source data for the figures in this study is available from Zenodo\cite{data}. \wtliwtli{DOI: 10.5281/zenodo.15859709}.

\section*{Acknowledgements}
We acknowledge Tong Jiang for helpful discussions.
This work is supported by National Natural Science Foundation of China (T2350009, 22433007),
the Young Elite Scientists Sponsorship Program by CAST (2023QNRC001),
University Development Fund (UDF01003789),
and the Shenzhen Science and Technology Program (No. KQTD20240729102028011). 
Shi-Xin Zhang acknowledges the support from a start-up grant at IOP-CAS.


\end{document}